# Performance of chemically modified reduced graphene oxide (CMrGO) in Electrodynamic Dust Shield (EDS) applications

Micah J. Schaible[a,b,*], Kristoffer G. Sjolund[c], Emily A. Ryan[d], Meisha L. Shofner[d], John R. Reynolds[a,d], Julie S. Linsey[b,c], and Thomas M. Orlando[a,b,e]

[a]School of Chemistry and Biochemistry, Georgia Institute of Technology
[b]Center for Space Technology and Research, Georgia Institute of Technology
[c]George W. Woodruff School of Mechanical Engineering
[d]School of Materials Science and Engineering, Georgia Institute of Technology
[e]School of Physics, Georgia Institute of Technology
*Corresponding author: mjschaible@gatech.edu

**Highlights:**

- Spray-coated EDS systems were produced using a conductive nanocomposite material
- 2-phase and 3-phase device configurations efficiently removed >80% of deposited dust
- The 2-phase devices were cleaned at ~50% lower voltage with illuminated with UV light
- Using a dielectric cap eliminates electrical discharges on the surface

**Abstract**

Electrodynamic Dust Shield (EDS) technology is a dust mitigation strategy that is commonly studied for applications such as photovoltaics or thermal radiators where soiling of the surfaces can reduce performance. The goal of the current work was to test the performance of a patterned nanocomposite EDS system produced through spray-coating and melt infiltration of chemically modified reduced graphene oxide (CMrGO) traces with thermoplastic high-density polyethylene (HDPE). The EDS performance was tested for a dusting of lunar regolith simulant under high vacuum conditions (~$10^{-6}$ Torr) using both 2-phase and 3-phase device configurations. Uncapped (bare) devices showed efficient dust removal at moderate voltages (1000 to 3000 V) for both 2-phase and 3-phase designs. Further tests carried out while illuminating the dust surface with a UV excimer lamp showed that the EDS voltage needed to reach the maximum cleanliness was reduced by almost 50% for the 2-phase devices (500 V minimum for rough and 1000 V for smooth), while the 3-phase devices were unaffected by the application of UV. However, the performance of the all uncapped devices degraded after several sequential tests due to erosion of the traces caused by electric discharges and dielectric breakdown. Capping the CMrGO traces with low-density polyethylene (LDPE) eliminated breakdown of the materials and device degradation, but larger voltages (3000 V) coupled with UV illumination were required to remove the grains from the capped devices.

**Keywords:**

Electrodynamic dust shield, nanocomposite, chemically modified reduced graphene oxide (CMrGO), Artemis, dust mitigation

## 1. Introduction

The return of manned missions to the Moon with the Artemis program will require novel solutions to the unique problems posed by lunar dust. The lunar surface is covered in an approximately meter thick layer of dust formed by billions of years of meteorite impacts which have pulverized the surface into a fine powder. The dust is readily disturbed by robotic and/or human activities in the lunar gravity environment, and the lack of atmosphere means that the grains can accumulate large electrostatic potentials. Electrostatically charged grains can strongly adhere to surfaces such as astronaut suits or solar panel surfaces, causing risk of reduced performance and failure.[1, 2] Subsequent tissue exposure in habitable environments can cause adverse health effects such as lung and eye irritation.[3, 4] In particular, the smallest size fraction (<5 μm) is known to pose the greatest health risk, and the smallest grains are also the most difficult to remove using current mitigation technologies. A diverse array of dust mitigation solutions is necessary to ensure the success of extended human exploration missions on the lunar surface.

Multiple dust mitigation technologies have been explored and developed over the last several decades in anticipation of a return to the Moon. These include fluid-based devices, mechanical methods, passive coatings, and electron bombardment.[5-8] One of the earliest technologies that was identified as a promising dust mitigation solution is Electrodynamic Dust Shielding, or EDS.[9, 10] EDS is an "active" (as opposed to passive) dust mitigation technique that uses alternating electric fields applied to one or more electrodes to remove dust from surfaces. Typical EDS devices include conducting electrode materials in thin-film form (e.g., indium-tin-oxide (ITO), carbon nanotubes, or silver nanowires) deposited on an insulating surface in a well-defined pattern, and the electrodes are typically capped with a dielectric material to eliminate shorting when operating at high voltages. On the lunar surface, EDS materials can be incorporated into surfaces that are expected to experience deleterious performance due to dust collection, with typical applications including solar panels,[11, 12] camera lenses and windows,[13] thermal radiators,[14] and spacesuit fabrics.[15, 16] Additional implementations for Earth-based solar panels are also being explored.[17, 18]

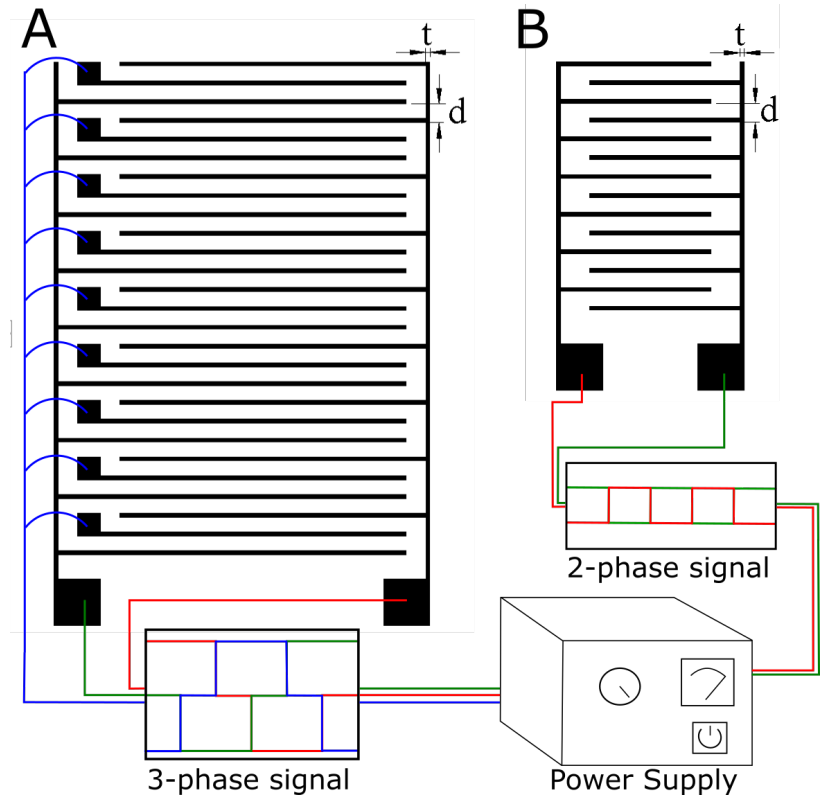

*Figure 1. Illustration of the EDS patterns used in these experiments and associated dimensions. The three-phase device is shown in (A) and the two-phase device is shown in (B).*

To initiate dust movement in EDS, a low-frequency, high-voltage AC signal is applied to the electrodes. Typical interdigitated device designs (shown in Figure 1) can be either 2-phase, consisting of two parallel electrodes, or 3-phase, consisting of three electrodes in either a rectangular or circular pattern.[19] The 2-phase devices can be run with one electrode grounded (1-phase operation) or with an AC signal applied to both electrodes (2-phase operation), while the 3-phase devices require a 3-phase signal. The applied fields set up a varying electric field above the device surface, where a standing wave potential is generated for 2-phase and a traveling wave for 3-phase devices. For the particles on the surface to be affected by the applied fields, they must themselves become electrically charged. In dielectrically capped devices, micro-discharges produced at the edges of the electrodes create numerous charged species, and the mobile electrons can quickly diffuse to the positively charged electrode while the positive ions migrate through the dielectric and can accumulate on grain surfaces.[18] Coulombic and dielectrophoretic forces from the EDS signals then mobilize the grains. Dust can accumulate additional charge through triboelectric charging, photoelectric charging, or plasma charging. While EDS has been shown to be effective for several applications and form factors, further development and combination with other mitigation solutions are needed for safe and reliable lunar surface operations.

In this work, we demonstrate the effectiveness of nanocomposite-based EDS systems, made by spray processing and melt infiltration of chemically modified reduced graphene oxide (CMrGO) and high-density polyethylene (HDPE),[20] in removing unsieved lunar highlands regolith simulant LHS-1 under high vacuum conditions. It is shown that the nanocomposite EDS devices can efficiently remove deposited dust from their surfaces, and that UV illumination decreases the voltage required for dust removal from 2-phase devices. This work specifically addresses EDS solutions for insulating, flexible polymeric surfaces, such as space-suits, inflatable habitats, and flexible robotics components, for which a limited number of EDS solutions exists.[16] Leveraging a thick, fully integrated nanocomposite layer as the electrically conductive traces permits the development mechanically robust, flexible EDS devices.

## 2. Experimental setup and methods

### 2.1 Device fabrication

EDS devices were fabricated by patterning a surface-localized nanocomposite layer using melt infiltration to flow polymer substrate material into spray-cast nanoparticle traces (see Figure 2). Labeled diagrams of the devices are shown in Figure 1. The electrode trace thickness ('t') was 0.5 mm and the inter-electrode spacings ('d') were 2 mm for all devices used. The spray casting and infiltration approach produced a device composed of only CMrGO, modified with linear dodecyl substitutes,[20] (Figure 2A) and the chosen thermoplastic substrate, HDPE. Surface-localized nanocomposite traces were made by dispersing CMrGO in chloroform ($CHCl_3$) at concentration of 10mg/ml and using a commercial airbrush (iwata Eclipse) and custom steel mask to pattern the CMrGO onto HDPE substrates (Figure 2B). After deposition, sprayed substrates were placed in a mold and static pressure and heat was applied to melt infiltrate the HDPE substrate into the CMrGO layer to create patterned nanocomposite regions fully integrated with the device surface (Figure 2C). Careful selection of the processing temperature, time, and pressure during infiltration allowed for direct control over the extent of infiltration and resulted in surface morphology which was varied from topographically rough and partially consolidated to smooth and fully consolidated, rough and smooth devices, respectively. After infiltration, devices were briefly sonicated in chloroform and wiped clean prior to use. This cleaning procedure removed any CMrGO particles that were not at least partially embedded in the HDPE matrix. Average sheet resistance measured using a 4-point line probe at five separate locations can be found in Table S1.

Three device configurations are reported here: rough, smooth, and capped. A lower processing temperature, 129°C, was used for the rough condition, limiting the extent of polymer infiltration, and a higher processing temperature, 139°C, was used for the smooth condition to fully infiltrate the CMrGO layer. The rough and smooth conditions were tested in their as-produced (uncapped) condition (Figure 2C). For the capped devices, a second low density polyethylene film (0.02-0.03mm, LDPE) was placed atop the as-sprayed device and the entire stack was subject to the rough condition melt infiltration treatment. By processing the thermoplastic capping layer at the same time as the substrate, the cap infiltrated the traces from the top side which resulted in a fully composite device composed of CMrGO nanocomposite 'wires' surrounded by polyethylene (Figure 2D). After processing, the cap layer was well-adhered to the HDPE surface and was unable to be separated from the device. The EDS devices used in the experiments described here were made using a black-colored HDPE (McMaster-Carr) surface to provide a high contrast with the light grey colored lunar simulant and improve the quality of the analysis images described below. The nanocomposite-based devices were compared to commercially produced printed circuit board (PCB) based devices that used the same 2-phase EDS pattern. In these devices, the traces were Cu instead of CMrGO, and the devices were coated with an unspecified dielectric epoxy instead of LDPE.

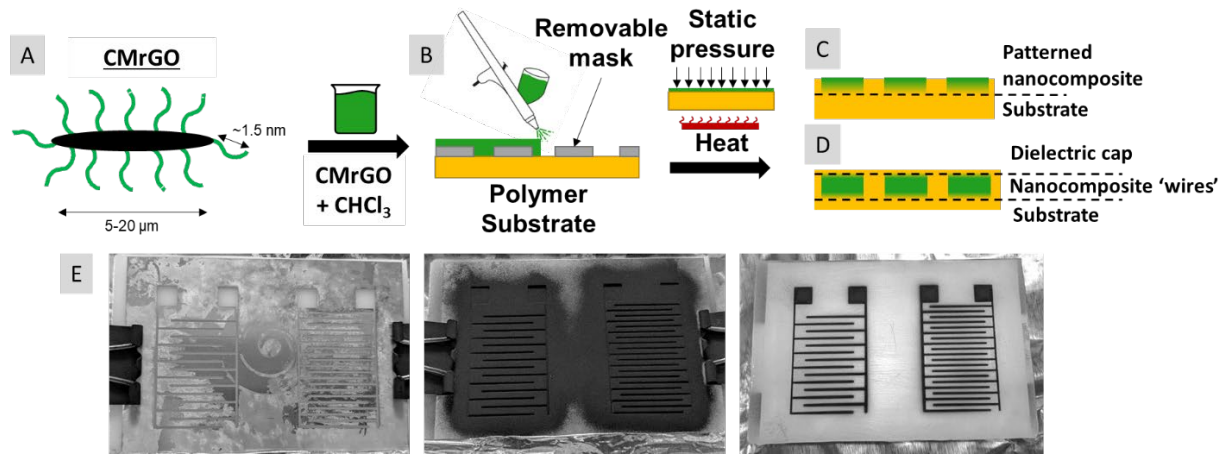

*Figure 2. (A) Reduced graphene oxide chemically modified with dodecyl substituents (CMrGO) was dispersed in CHCl$_3$ and (B) sprayed over a removable steel mask. (C,D) Sprayed substrates were then subject to heat and pressure to create an infiltrated surface-localized nanocomposite device. (E) Masked substrate before spraying (left), masked substrate sprayed with CMrGO (middle), patterned EDS devices on HDPE surface (right). White HDPE substrates are shown in these images to enhance visual contrast for photographs.*

## 2.2 Lunar simulant exposure

To test the CMrGO EDS devices, the conductive traces were connected to external power supplies to generate the appropriate AC signals using Kapton-coated copper wires affixed to the integrated connection pads with stainless steel screws. Each sample was mounted horizontally in a vacuum chamber and subsequently coated to ~100% coverage with an even layer of lunar highlands regolith simulant LHS-1 (Exolith lab, University of Central Florida), as shown in Figure 3A. The particle size distribution of LHS-1, determined from optical microscopy images, ranges from <1 to >500 μm, with an effective diameter of ~2 μm and a mean circularity of ~0.55.[21] A Ni wire screen was used to evenly deposit the dust and remove the largest size fraction of the dust (> 0.3 mm), and ~0.1 g was deposited on the 2-phase devices and ~0.3 g on the 3-phase devices. The precise weight of the deposited dust could not easily be determined due to some of the dust falling onto the chamber walls during application. The EDS signal was applied through an MHV vacuum flange located below the device, and the chamber was held at high vacuum pressures (~$10^{-7}$ Torr), although the pressure increased (~$10^{-6}$ Torr) during the experiments.

The EDS signals were supplied to each electrode by ambipolar power supply modules (Matsusada, AS-1.5B2(A)) that provided a maximum output of 3000 V peak-to-peak (±1500 V) for arbitrary waveforms. The power supply modules were connected to a custom LabView control program that allowed independent control over the waveforms and phase offsets sent to each electrode. All experiments described here used a 10 Hz square wave for the applied EDS signal. A 180° phase shift was used for 2-phase devices and 120° phase shift for the 3-phase. To determine the effects of photoelectric charging on the EDS efficiency, a 172 nm UV excimer lamp (Ushio, UXFL-UST10L70-01A) was mounted inside the chamber about 5 cm above and slightly to the side of the EDS devices (Figure 3A). The UV lamp had a reference UV intensity of 2.1 mW/cm$^2$ at 5mm, and the cylindrically symmetric active length of 100 mm was roughly centered on the EDS device such that the entire surface was illuminated uniformly. For experiments including UV exposure, the UV was turned on with the EDS signal, and then cycled off and on at 5 second intervals for the duration of the experiments (~5 min). Because radiative heating and cooling dominate in the vacuum environment, the sample surface warmed to ~65 °C after >1 min of continuous exposure (as measured by a K-type thermocouple attached to the sample surface). Pulsing the lamp led to minimal surface heating (<30 °C).

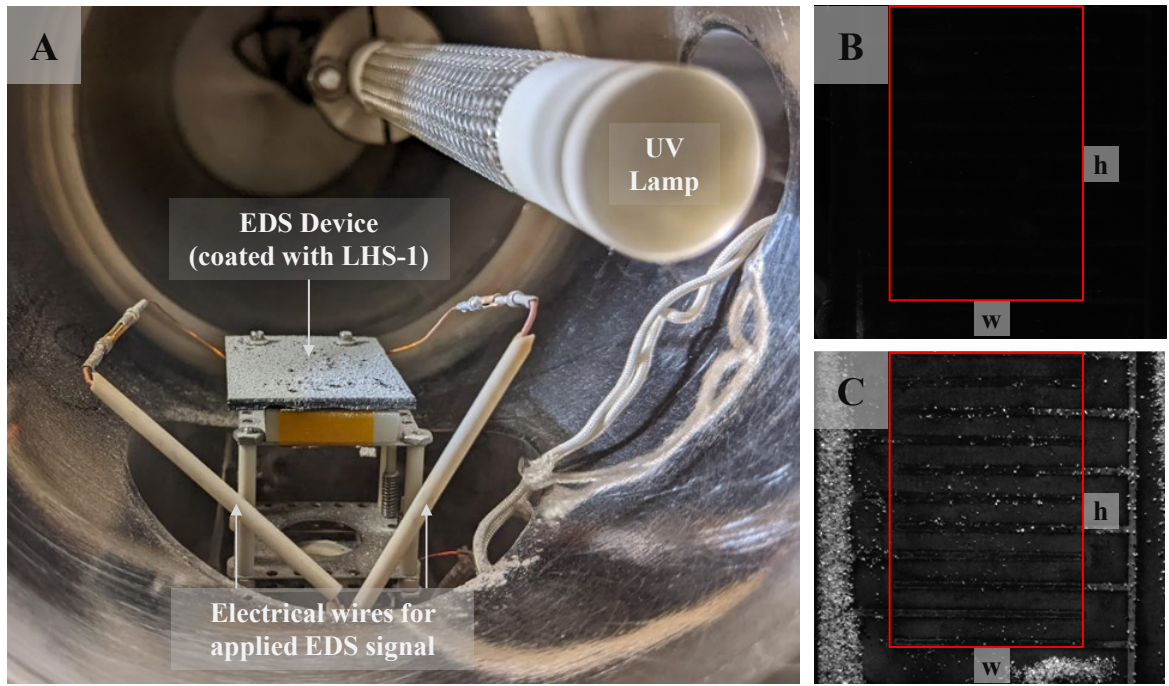

*Figure 3. (A) Experimental system with 2-phase sample installed and coated with a layer of dust. The UV excimer lamp is ~5 cm above the device surface. (B) Camera images of the device surface prior to coating with dust and (C) after running the EDS for ~5 min. The area of analysis, indicated by the red box, had a width of about 13mm and a height of about 21mm.*

To determine the dust removal efficiency, time-lapse images were captured at 1 sec intervals while the the EDS was active. The devices were illuminated with a collimated, 455 nm LED light source (Thor Labs, M455L4) mounted outside the experimental chamber to provide uniform lighting across the entire surface. The intensity of the LED did not cause any observable heating, and the brightness was sufficient such that no change in illumination conditions could be detected during UV lamp pulsing. Photos of the bare and cleaned devices (see Figure 3B,C, respectively) were taken from directly above using a monochromatic camera (Manta, 146B GigE) with a 25 mm focal length lens (Navitar). The surface cleanliness was determined using the ImageJ software to determine the percentage of the surface covered by dust using a brightness thresholding technique.[22] The area of analysis of the 2-phase devices was restricted to only the area where the electrodes overlapped (see Figure 3C), while for the larger 3-phase devices the entire captured image was analyzed. A binary mask of each image was created using a threshold brightness value, and pixels with brightness values higher than the threshold were interpreted as dust (uncleaned) while values below the threshold were considered background (cleaned). The percentage of cleaned pixels for each frame was then calculated, this value was used as the cleanliness for each sample at the given point in time. After each experiment, a Kimwipe dampened with methanol was used to wipe any residual dust from the surface of devices. Once completely dry, the surface was re-dusted as described above for subsequent testing.

*2.3 Pre- and post-exposure materials analysis*

Surface resistance measurements of the surface localized nanocomposite materials were made using a four-point line probe with pins spaced at 10 mm (Signatone SP4) and a Keithley 2400 source meter. Resistance measurements were made on 26 × 13 mm regions produced under the sample infiltration conditions on the same substrates. Samples were measured at five locations, an average value and standard deviation for both smooth and rough conditions are reported in the Supplementary Information, Table S1.

For morphological characterization, a Phenom XL G2 scanning electron microscope (SEM) was used to obtain images at a number of different magnifications. Before obtaining SEM images, the samples were rinsed with isopropanol (IPA) to remove any loose dust and then blown off with dry compressed air or argon. Sections were then cut from the devices and were sized to include at least two traces. Cut sections were mounted to SEM stubs and gold-coated to allow for imaging of the non-conductive inter-trace regions as well as any non-conducting adhered dust. Any dust visible in SEM images is well adhered to the surfaces and was not removed during rinsing.

## 3. Data and results

*3.1 Dust clearance for as-produced CMrGO devices*

Rough and smooth CMrGO devices were tested under high vacuum conditions with a deposited dust layer that just obscures the underlying EDS device. Although the thickness of the dust layer was not directly measured, it was clear from the images that the highest points on the surface were due to the largest (and least abundant) dust grains, such that the thickness could be estimated as <0.5 mm. For the rough devices, at EDS voltages above about 1000 V, small discharges or sparks were observed on the edges of most of the traces for several seconds after EDS initiation, even in the absence of dust. The number and intensity of the discharges increased with increasing voltage, and the presence of dust appeared to density of discharges visible immediately after EDS was initiated. In both bare and dust-coated conditions, the discharges typically subsided within 10-20 seconds, although some spots on the traces would continue to spark throughout the experiment.

The rough devices cleared the dust at significantly lower voltages than the smooth devices, 1000 V vs 2000 V, respectively, and were routinely able to achieve a higher percent cleanliness as shown in Figure 4A. For both device configurations, most of the dust was cleared within a few seconds of initiating the EDS, and a steady state cleanliness was typically reached in 10-20 seconds. The EDS cleaning behavior was repeatable for similarly prepared devices, and the cleanliness could be obtained for at least three sequential re-dusting and activation tests. However, there was an eventual decrease in the grain removal efficiency with additional testing for both rough and smooth devices. This led to a more gradual clearing of the dust grains from the surface and a reduction in the maximum % cleanliness achieved for the same EDS testing parameters. The differences in the initial cleaning efficiencies immediately after the EDS is turned on is indicted by the large initial confidence interval for the smooth device (Figure 4A); no smooth device ever reached greater than 85% cleanliness, and the larger uncertainties in the 95% confidence intervals for the first 20 sec are due to different rates of cleaning. For the rough devices, significant degradation in the performance was typically evident by the fifth test. This effect was even more pronounced in the UV tests described further in §3.2. A visual demonstration of the progression in surface cleanliness is shown in Figure 4B, where the lower panel indicate that residual dust tended to accumulate on the conductive traces.

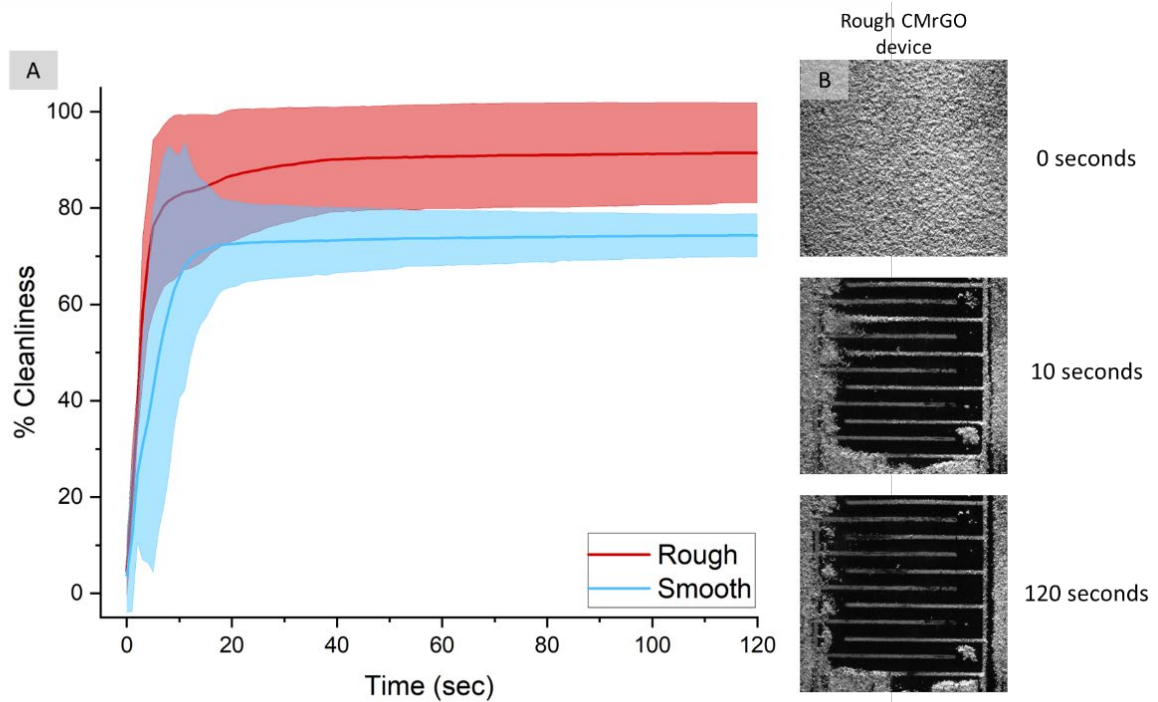

*Figure 4. (A) Surface cleanliness achieved for the CM-rGO EDS devices as a function of time using 1000V peak-to-peak applied to the rough (red) and 2000V peak-to-peak applied to the smooth (blue). The mean cleanliness with 95% confidence intervals were determined from 3 experiments for each condition. (B) Images of the rough sample during one of the experiments. The top image was the initial condition, the middle image was ten (10) seconds into the experiment, and the bottom image was the was one-hundred-and-twenty (120) seconds into the experiment after a steady state had been reached.*

SEM images were collected after device testing to evaluate the underlying mechanism of sparking and device performance degradation observed during testing. SEM images in Figure S1A,D show that rough device surfaces initially had a partially porous structure which consisted of CMrGO platelet structures partially embedded in HDPE and an abundance of micron to few micron sized pores, which were on the same size order as the majority of the LHS-1 simulant particles [21]. An image of the as-sprayed CMrGO particles on a segment of carbon tape is shown in Figure S2. For the device surfaces that were exposed to dust and EDS voltages, shown in Figure S1B,C and E,F, CMrGO platelet edges appeared rougher, and dust appeared to be strongly associated with the crevice regions of the devices, seen as bright, angular objects in the SEM images (indicated by red arrows). These changes in the surface morphology of the CMrGO trace regions support the concept that micro-discharges near the rough edges of CMrGO platelets contributed to the dust removal from the surface.

The unexposed smooth devices in Figure S3A,D showed smooth surfaces with very little porosity or surface texture in the nanocomposite trace region, which covers roughly the center 500μm of the image, due to full infiltration of the CMrGO layer by HDPE. In Figure S3B,E significant degradation was noted at the trace edges. Deep voids, rough edges, and exposed CMrGO particles indicated that erosion of material occurred, likely due to micro-discharges and/or dielectric breakdown. The higher operating voltages needed for clearance on the smooth devices may be required to promote micro-discharges when additional HDPE is present at the surface in fully infiltrated devices. Like the rough devices, smooth devices degrade with sequential tests, although similar device performance could be maintained for as many as 10 sequential tests before the degradation became apparent. The slowed degradation rate may be due to the CMrGO particles being fully embedded in the substrate, thus providing greater resistance to erosion of the material during micro-discharge and dielectric breakdown events. Regions where the highest density of discharges were observed during EDS operation were also those that experienced that largest amount of degradation (evidence of micro-discharges along the edges of the CMrGO traces and larger breakdowns that spanned entire traces, shown in Figure S4). Extended use of the devices resulted in large amounts of ingrained dust and sometimes complete failure of individual traces due to continued erosion and material breakdown. As discussed further in §3.2, exposing the surfaces to UV during EDS operation led to increased degradation of the device performance.

The smooth 3-phase device tested showed no improvement in performance over the 2-phase devices, and larger applied voltages were required to remove grains from both rough (1500 V) and smooth (3000 V) devices. Because discharges were also observed on the edges of the rough 3-phase device traces for applied voltages of >1000 V and this led to rapid degradation, data is only shown for the smooth 3-phase devices. Although the maximum removal efficiency was similar, the as-produced 2-phase devices were observed to approach their maximum cleanliness within 10-20 seconds (see Figure 4A), while the 3-phase devices typically required >50 seconds to reach the maximum dust removal, even in the very first device test after production (Figure 5A). The vacuum chamber pressure was also observed to have a notable effect on the efficiency of the smooth 3-phase devices, which was not the case for the 2-phase devices. Experiments run after pumping the chamber for several hours showed significantly worse performance than the devices that were pumped overnight (see Figure 5A). The degradation of performance in the smooth 3-phase device due to sequential activation and re-dusting was lower than the smooth 2-phase device, despite higher operating voltages. Figure 5B shows a series of cleanliness images for the smooth three-phase device at $10^{-7}$ Torr operating conditions. Similar to the 2-phase devices, the dust segregated to the conductive trace regions.

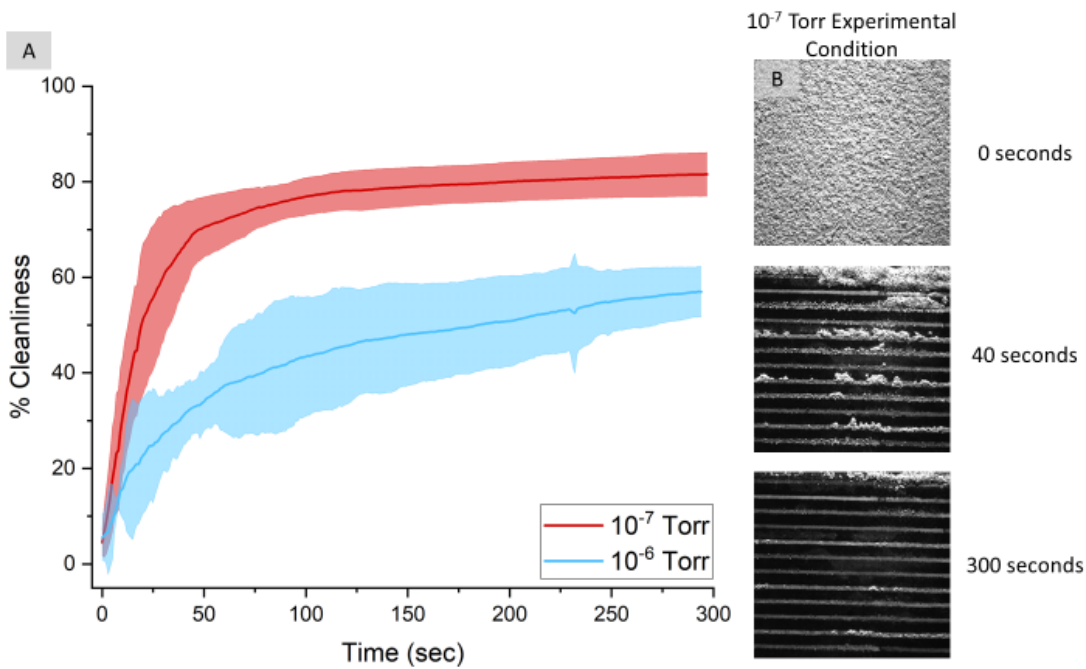

*Figure 5. (A) Surface cleanliness versus time achieved using the 3-phase smooth CM-rGO EDS devices after pumping the vacuum chamber for about four (4) and twenty-four (24) hours (blue "$10^{-6}$ Torr" and red "$10^{-7}$ Torr", respectively). (B) Images of the $10^{-7}$ Torr sample during one of the experiments. The top image was the initial condition, the middle image was forty (40) seconds into the experiment, and the bottom image was the was three-hundred (300) seconds into the experiment when the experiment had ceased.*

*3.2 Dust clearance for CMrGO devices illuminated by UV light*

Because the EDS process relies on excess charge on the surfaces of the dust grains, UV illumination was considered as a method to enhance grain charging and clearance. UV photons eject electrons from the surfaces of the grains and the substrate, and these excess electrons can accumulate in "patches" on the surfaces of the dust particles.[23] Experiments which included UV exposure showed similar dust clearance behavior to the non-UV tests where the grains were rapidly removed from the surfaces and an average cleanliness of ~90% was achieved for the first test after device production. However, the maximum grain removal was achieved at applied voltages that were roughly half of those required for similar removal efficiencies in the non-UV case; the dust could be efficiently removed with as little as 800 V peak to peak for the rough sample and 1000 V for the smooth. Leaving the UV on resulted in a steady state cleanliness being achieved ~20sec after starting the experiment, while cycling the UV on and off in five second intervals was observed to promote more dynamic dust motion and improved the overall dust clearance at lower applied EDS voltages.

It was observed that, while the UV was on, there was an increase in the frequency and abundance of discharges for similar operation voltages as in the non-UV experiments. Further, subsequent activation and re-dusting tests showed that UV illumination caused the devices to degrade more rapidly, but in the same manner as the non-UV tested devices, as shown by the erosion of the device surfaces in Figure S1C,F and S3C,F. The series of runs shown in Figure 6 for both the rough and the smooth samples show a clear deterioration in performance after the first test. This trend was not observed for the non-UV tests presented above in which variable cleanliness results had no clear trend associated with activation and re-dusting. In the UV tests for the rough sample, test 1 showed rapid and near complete (>95%) removal of the dust from the device surface at only 800 V, but by the fourth test the rate of cleaning was noticeably delayed, as shown in Figure 6A, and the maximum cleanliness reached during the experiment was only about 85%. Subsequent testing on these devices with UV continued to degrade the performance. A similar trend was noted in the smooth devices tested with UV, where initially similar cleanliness levels to the non-UV tests were achieved, but subsequent tests slower cleaning and reduced ultimate cleanliness (Figure 6B).

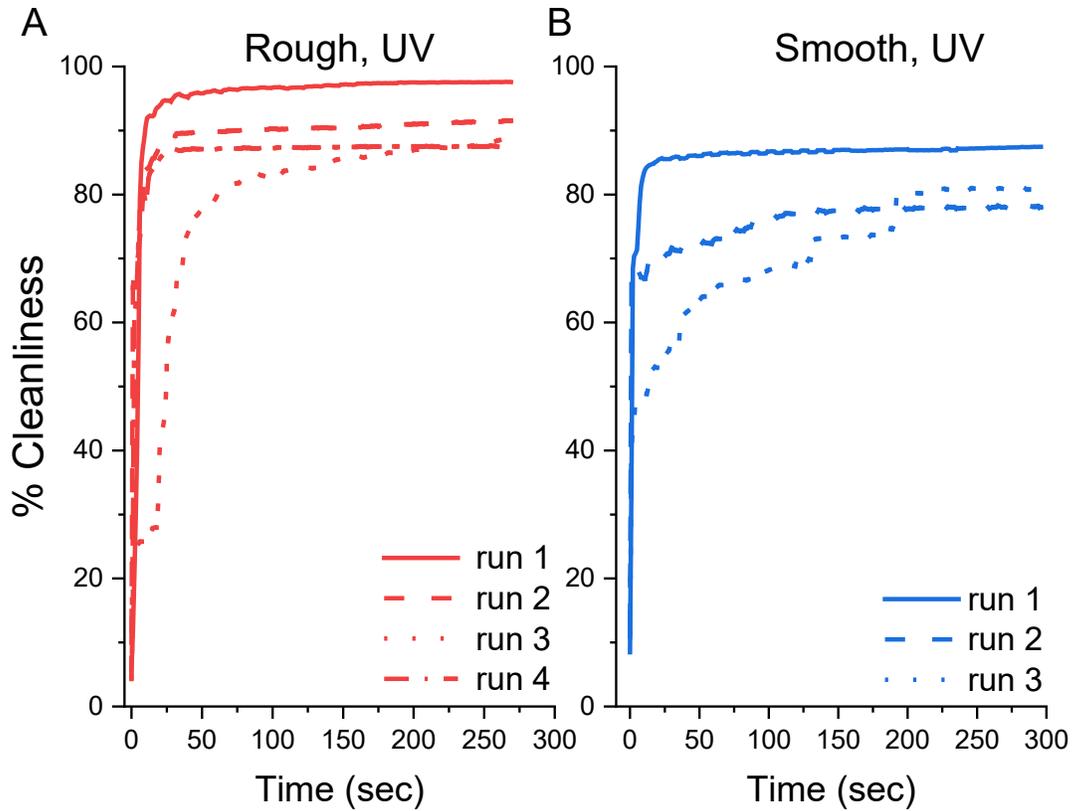

*Figure 6. Line plots depicting the % cleanliness achieved using (A) 800V peak-to-peak applied to the rough configuration and (B) 1000V peak-to-peak applied to the smooth configuration while illuminating the dust surface with UV light.*

*3.3 Dust clearance for CMrGO-PE capped devices and comparison to PCB*

Due to the erosion observed in the uncapped rough and smooth devices with and without UV, a capped device configuration was considered. Many prior examples of EDS systems for lunar and Martian environments include a dielectric capping layer as a component of the design to isolate the active traces from the dust [9-14]. In addition to electrically isolating the traces, the capping layer served as a physical barrier to prevent erosion due to micro-discharges and dielectric breakdown in the composite trace regions,

For this study, a second thermoplastic layer was chosen as the capping layer maintain a fully-composite device architecture that can be assembled without the need for adhesives or other secondary materials.

The EDS experiments that used capped devices showed reduced performance compared to the bare devices despite higher operating voltages. The capped devices were all the 2-phase design. For all capped devices, even at the maximum voltage available (3000 V), exposure of the dusted devices to UV was required to observe any significant amount of grain motion or surface cleaning. In the absence of UV, only a very small number of grains were observed to leave the surface and there was no significant effect on the coverage. However, for experiments including UV, the capped devices were able to effectively and repeatably clean the surface. No discharges were visible during the experiments for any of the capped devices, and no evidence of degradation after experimentation was observed in SEM images. However, like the uncapped devices, the grains tended to accumulate over the traces and clear completely between traces. Additionally, there was some evidence of grain alignment along electric field lines at the ends of traces. Small groupings of grains formed <0.1 mm wide lines perpendicular to the electrodes aligned with the expected electric fields.

The percent removal for the low-density polyethylene (LDPE) and PCB capped devices is shown in Figure 7. The LDPE capped device showed more variable cleaning performance than the PCB device, but there was no trend in the cleanliness indicating degradation of the device as seen in the uncapped conditions exposed to UV. Variability of the cleaning performance of the LDPE capped device manifested as different final cleanliness values with no significant change in the rate of cleaning during subsequent tests. The PCB device attained consistent cleanliness values; however, they did not perform as well as the uncapped CMrGO devices, despite the larger applied EDS potentials.

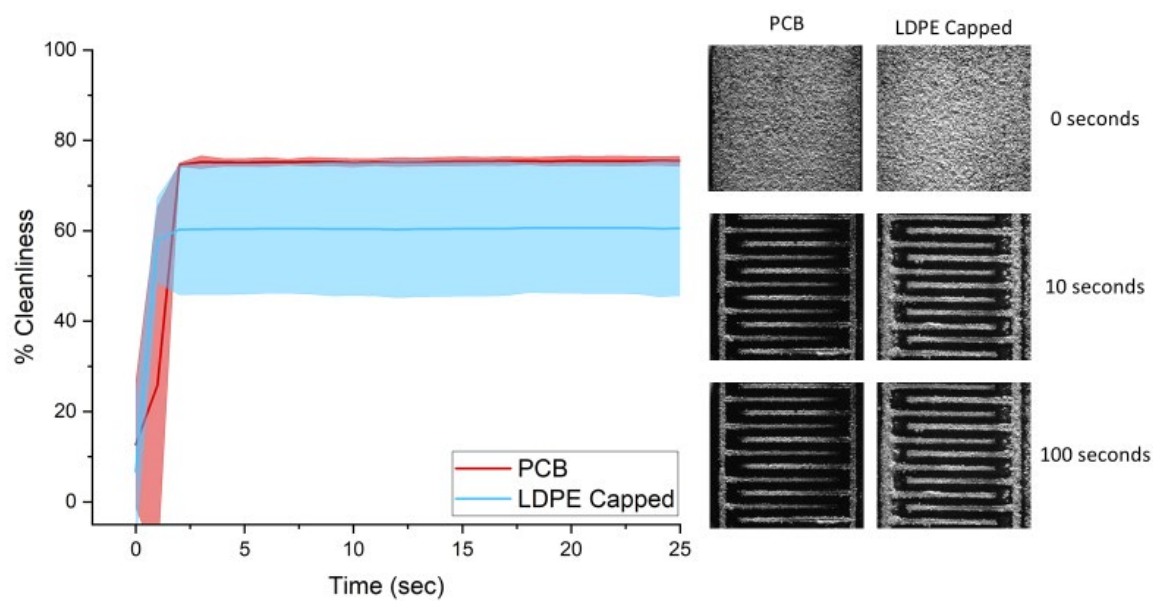

*Figure 7. Surface cleanliness versus time achieved using the PCB EDS chip (red) and the LDPE capped CM-rGO EDS chip (blue) with UV photoelectric charging. The mean cleanliness with 95% confidence intervals were determined from 3 experiments for each condition.*

## 4. Discussion

The experimental results described above clearly demonstrate that the newly developed conducting CMrGO nanocomposite material can be successfully used in EDS device applications. Even after coating the devices with a multiple grain thickness dust layer, cleaning from around the electrodes began as soon as the EDS was activated. Similarly, exposing the dust surface to UV light produced an immediate effect for the 2-phase devices. In all cases, the spaces in-between the traces were seen to clear first, and the dust preferentially segregated to the traces that then cleared more gradually. The tests described in §3.1 and §3.2 were carried out on bare (uncapped) devices, and post testing analysis showed that the traces were clearly deteriorated where discharges were most concentrated. After several tests the surface was noticeably eroded as shown in Figures S1 and S3, with most damage occurring mostly along the edges of the traces.

Discharges were created when high voltages were applied to the CMrGO electrodes, likely due to breakdown occurring at CMrGO-CMrGO junctions and/or the electrodes arcing to charged dust grains. Due to the melt infiltration construction, HDPE is likely present at most particle-particle junctions, and micro-discharges in these regions likely result in dielectric breakdown and material degradation. The sharp edges of the CMrGO platelets allow for very high fields to build up, and when applying the EDS potentials, this leads to field emission and electron avalanche discharging that creates additional charge carriers,[24, 25] thus increasing the EDS forces felt by the dust grains, as well as explosively disrupting the surface. The field strengths and field emission effects will be strongest at the junctions where conducting particles are separated by a thin insulating layer. This effect was especially strong in the rough CMrGO devices which had a more porous structure, more exposed CMrGO edges, and less HDPE surrounding the particles providing resistance to erosion. The increased frequency of discharges on rough devices likely contributed to the increased cleaning efficiency at lower applied voltages for than the smooth devices, despite a similar trace resistance as the smooth devices.

Illuminating the devices with UV light lowered the voltage required to remove the dust from the surface by almost 50% for the 2-phase devices. The increased efficiency for the 2-phase devices can be explained by photoelectric charging caused by the UV photons. Photoelectrons emitted from both the grains and the substrate can accumulate on the insulating grains surfaces, thus increasing the forces felt from the EDS potentials. Further, photoelectrons can lead to increased discharges from the CMrGO materials, and an increased number of discharges were overserved at lower applied voltages in the presence of UV. The increase in discharges was accompanied by more rapid device degradation, and clear performance degradation was obvious in the UV tests after only one run, whereas in the absence of UV the devices typically operated at a similar efficiency for several runs. Capping the devices prevented these micro-discharges from causing explosive erosion by encapsulating the conducting traces while maintaining the percolated CMrGO network. For the capped devices, no degradation was seen in the device performance or in post-testing analysis. Further work is needed to understand if internal porosity is generated in the capped devices or the encapsulated traces are fully protected from degradation by capping.

Interestingly, the 3-phase devices showed remarkably different behavior than the 2-phase devices. UV illumination did not appear to have any effect on the efficiency of dust removal for the 3-phase devices. While this difference was likely due to differences in how the grains accumulate charge in the two different configurations, there is not currently a precise description mechanism for lack of an effect in the 3-phase devices. Further, it was observed that the performance of the 3-phase devices was directly affected by the amount of time that the dust was exposed to the high vacuum conditions of the experimental chambers. When the dust is exposed to vacuum, atmospheric water vapor and other adventitious species will slowly desorb from the surfaces. As these species are pumped away and the underlying mineral surface sites are exposed, the intergrain cohesive forces may be reduced and/or the grains may more efficiently accumulate charge. Further experiments are necessary to determine the precise relationship between EDS efficiency and background gas pressure for the devices described here.

Leveraging surface-localized, or fully embedded, CMrGO percolated networks on flexible substrates opens the possibility for flexible and conformal, integrated EDS systems. The orientation agnostic processing method allows for device fabrication in which the substrate and thermoplastic dielectric cap are both integrated during the same infiltration step to produce fully composite devices. Good dispersibility of CMrGO in a variety of solvent systems also means that the spray processed inks used in this study could be adapted to direct write inks to produce in-space manufacturing (ISM) compatible device fabrication schemes. Inkjet, direct write, and fused filament fabrication have previously been proven as ISM technologies and have planned integration into Artemis as well as future shuttle and Mars missions.[26] Additionally, the melt infiltration processes used here is not expected to be disturbed by the low gravity environment as the infiltration process relies primarily on capillary action and wetting phenomena to backfill the nanocomposite traces, and not on gravitationally assisted flow.[27] Finally, because thermoplastics can be re-flowed several times, this ink deposition and infiltration method provides a pathway to the repairability of damaged surfaces or components by simply re-coating and re-infiltrating the thermoplastic surface.

## 5. Conclusions

It is found that the CMrGO EDS devices operate well in high vacuum conditions and can remove >90% of the dust under typical voltage conditions (1000 V) or under reduced voltage conditions when incorporating UV exposure (800 V). However, the material degrades after several runs, and numerous discharges are seen on the surface during operation. The degradation in performance is more notable in the devices exposed to UV, and post-testing analysis showed that the explosive discharges removed significant amounts of the CMrGO material from the deposited traces and that small dust grains were able to infiltrate far below the surface. Placing a ~0.02 mm cap on the CMrGO EDS devices eliminated the discharges and material breakdown but required a higher voltage to remove the dust (3000V + UV exposure). Future work will explore alternative capping materials, including nanostructured materials to reduce surface adhesion effects and potentially improve the cleaning performance of devices for the smallest size grain fractions, and additional mitigation strategies such as vibrational tribocharging, and electron bombardment.

**Acknowledgements:**

The authors would like to thank GT School of Chemistry and Biochemistry support engineer Richard Bedell for his assistance with several of the lab instruments used in this work. The study was directly supported by the NASA Solar System Exploration Research Virtual Institute (SSERVI) under Cooperative Agreement #NNA17BF68A (REVEALS). Funding was also received from the 2021 NASA's BIG Idea Challenge as a part of team Shoot for the Moon from Georgia Institute of Technology. SEM images were collected in Georgia Tech's Materials Innovation and Learning Laboratory, an uncommon "make and measure" space committed to elevating undergraduate research in materials science, supported by Georgia Tech.